\documentclass[aps,prx,reprint,superscriptaddress,nofootinbib]{revtex4-2}

\usepackage{amsmath,graphicx,color}
\usepackage{bm}
\usepackage{amsfonts}
\usepackage{amssymb}
\usepackage{bbold}
\usepackage[colorlinks=true,linkcolor=blue,citecolor=blue,urlcolor=blue]{hyperref}
\usepackage[all]{hypcap}
\usepackage{graphicx}

\newcommand{\gin}[0]{{\gamma_\mathrm{in}}}
\newcommand{\gex}[0]{{\gamma_\mathrm{ex}}}

\newcommand{\vect}[1]{\bm{#1}}
\usepackage{verbatim}

\newcommand{\mc}{\mathcal}

\begin{document}

\title{Topological phases in discrete stochastic systems}

\author{Jaime Agudo-Canalejo}
\email{j.agudo-canalejo@ucl.ac.uk}
\affiliation{Department of Physics and Astronomy, University College London, London WC1E 6BT, United Kingdom}

\author{Evelyn Tang}
\email{e.tang@rice.edu}
\affiliation{Physics and Astronomy Department and Center for Theoretical Biological Physics, Rice University, USA}


\begin{abstract}
Topological invariants have proved useful for analyzing emergent function as they characterize a property of the entire system, and are insensitive to local details, disorder, and noise. They support boundary states, which reduce the system response to a lower dimensional space and, in 2D systems, offer a mechanism for the emergence of global cycles within a large phase space. Topological invariants have been heavily studied in quantum electronic systems and have been observed in other classical platforms such as mechanical lattices. However, this framework largely describes equilibrium systems within an ordered crystalline lattice, whereas biological systems are often strongly non-equilibrium with stochastic components. We review recent developments in topological states in discrete stochastic models in 1D and 2D systems, and initial progress in identifying testable signature of topological states in molecular systems and ecology. These models further provide simple principles for targeted dynamics in synthetic systems and in the engineering of reconfigurable materials. Lastly, we describe novel theoretical properties of these systems such as the necessity for non-Hermiticity in permitting edge states, as well as new analytical tools to reveal these properties. The emerging developments shed light on fundamental principles for non-equilibrium systems and topological protection enabling robust biological function. 
\end{abstract}

\maketitle

 \section{Introduction}
 
Understanding how the underlying components of proteins, RNAs, metabolites, and other biologically relevant molecules give rise to biological function would enable precise targeting of interventions to enhance health. However, we still do not have a good theory for how structure leads to function in these complex and noisy systems. As just one example, despite successful sequencing of the human genome, it remains challenging to predict the behavior of resulting proteins and macromolecules due to the large space of possible configurations and reactions causing transitions between them \cite{Ross2009ComplexBiology,Ashkenasy2017SystemsChemistry}. This large phase space of possible configurations also renders exhaustive searches using other approaches like experiment or numerical simulation unfeasible \cite{Dewyer2018MethodsSystems}, underscoring the need for simple conceptual methods to provide insight \cite{Davies2020DoesMatter,Winfree1980TheTime,Gao2015OnNeuroscience}. This review addresses how, drawing on recent theoretical advances in topology, molecular biology, and non-equilibrium statistical physics, the study of topological states in discrete stochastic models provides a theoretical approach to predict robust and dynamic function in living systems.

{\color{black}

In this context, topology refers to the study of global features of the system's eigenvalues and eigenfunctions, particularly in momentum (Fourier) space, that remain unchanged upon continuous deformation. The origins of topological classification stem from random matrix theory \cite{Mehta1967RandomLevels,Schnyder2008ClassificationDimensions,Ryu2010TopologicalHierarchy}. Topological invariants such as winding numbers can be defined that signal topological changes in these mathematical objects, which in turn manifest as phase transitions in the physical system, most famously leading to the emergence of boundary responses that are robust to generic or random perturbations, also known as edge states \cite{Moore2010TheInsulators,Hasan2010Colloquium:Insulators}, see Fig.~\ref{extfig:schematic}. 
Such topological invariants have proved useful for analyzing emergent function as they characterize a property of the entire system \cite{Moore2010TheInsulators,Chiu2016ClassificationSymmetries,Stormer1999TheEffect}.

\begin{figure*}
      \includegraphics[width=0.7\linewidth]{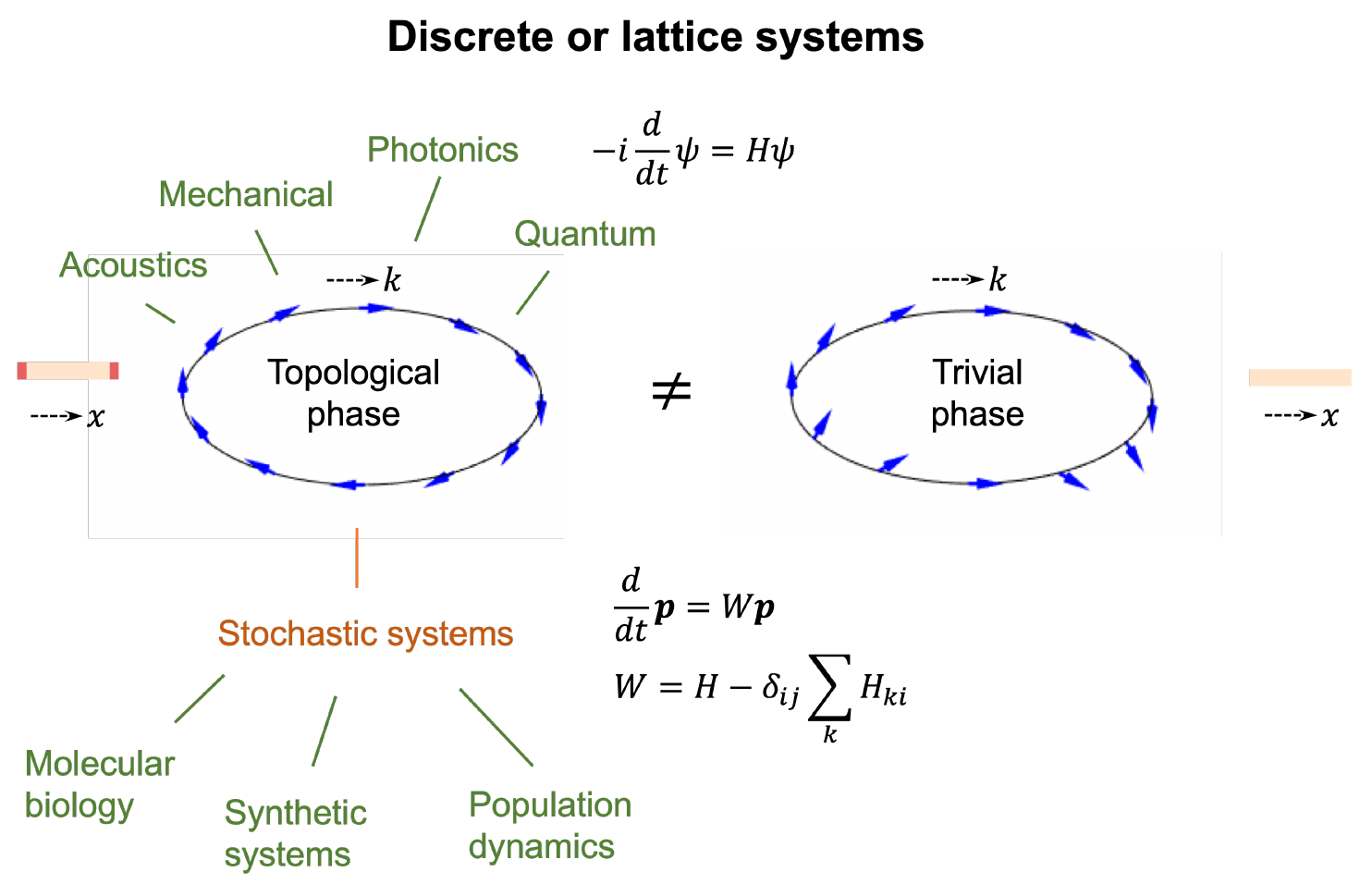}
      \caption{Topological invariants signal the emergence of edge states across diverse platforms. We depict an example of a topological invariant: a winding number. This can be defined in Fourier space, since most topological models are composed of repeated motifs (unit cells). The Fourier domain is periodic and shown as black circles. Transformed eigenvectors can have a phase (shown as blue arrows) that can wind by $2\pi$ as we move along the domain \cite{Obana2019TopologicalModel}.  \textit{Left}: The eigenvector phase winds around the domain and thus has a nonzero winding number. This indicates a topological phase hence will correspond to the presence of localized states in the real system on its edges, shown as red dots at the ends of a beige line.  \textit{Right}: In contrast, here the eigenvector phase fluctuates but does not wind by $2\pi$, so this trivial phase will correspond to an absence of edge states in the real system (beige line). Topological models have been developed in a variety of systems. \textit{Top}: Models were first developed in systems that can be described by the Schr{\"o}dinger equation or mapped onto it,  including quantum electrons \cite{Moore2010TheInsulators,Hasan2010Colloquium:Insulators}, mechanical lattices \cite{Mao2018MaxwellMechanics,Pedro2019TopologicalInteractions,Kane2014}, photonics \cite{Xiao2014SurfaceSystems,Benalcazar2017QuantizedInsulators,Yuce2019TopologicalModel,Pocock2018TopologicalEffects}, and acoustics \cite{Ni_2017,Liu2017PseudospinsLattice}. \textit{Bottom}: Discrete classical systems obey other dynamical equations such as the master equation, with various applications from molecular biology to synthetic systems \cite{Tang2021TopologySystems,Murugan2017TopologicallySystems,DasbiswasE9031}, or the Lotka-Volterra equation in the case of population dynamics models \cite{Knebel2020TopologicalCycles,Yoshida2021ChiralCycles}.\label{extfig:schematic}}
 \end{figure*}

Traditionally having been used in the quantum condensed matter field, the framework of topological invariants can identify the global state of the system and its associated response, for large classes of different systems \cite{Moore2010TheInsulators,Tang2014Strain-inducedInsulators,Weis2011MetrologyEffect,Cohen2019}. In quantum electronic systems \cite{Moore2010TheInsulators,Chiu2016ClassificationSymmetries,Tang2014Strain-inducedInsulators,Tang2011High-temperatureStates,Obana2019TopologicalModel,Lieu2018TopologicalModel,Su1979SolitonsPolyacetylene}, most notably, this framework enabled the discovery of dissipationless edge states that facilitate precise measurements enabling new standards in metrology \cite{Weis2011MetrologyEffect}. This framework has since been extended to various classical systems, from mechanical lattices \cite{Mao2018MaxwellMechanics,Pedro2019TopologicalInteractions,Kane2014} to photonics \cite{Xiao2014SurfaceSystems,Benalcazar2017QuantizedInsulators,Yuce2019TopologicalModel,Pocock2018TopologicalEffects}, acoustics \cite{Ni_2017,Liu2017PseudospinsLattice}, and active matter \cite{Tang2021TopologySystems,DasbiswasE9031,Murugan2017TopologicallySystems,Knebel2020TopologicalCycles,Yoshida2021ChiralCycles}. Regarding terminology, we must note that the Fourier-space, global topological phenomena described in this review are distinct from  real-space, local topological phenomena such as  knots in DNA \cite{Darcy2021DNALinks} and orientational defects in liquid crystals and biological tissues \cite{Ardaseva2022TopologicalMatter,Shankar2022TopologicalMatter}.

Recently, new applications of topological tools to non-equilibrium stochastic systems have been demonstrated, and shown to identify localization \cite{Murugan2017TopologicallySystems,DasbiswasE9031} and currents \cite{Tang2021TopologySystems} on the boundaries of an abstract reaction or configuration space. These systems are often composed of repeated motifs (unit cells) that allow for the flexible design of lattices with edge states, see Fig.~\ref{extfig:models}. The edge states manifest as distinct localization or dynamics on a lower dimensional submanifold of the larger system, without requiring fine-tuning or over-parametrization for this specialized behavior.  Powerfully, this edge response is insensitive to local details, disorder, and noise, and can describe robust biological oscillations as global cycles around the boundaries of a large two-dimensional configuration space.  They have been used to model dimensional reduction in the circadian rhythm \cite{Tang2021TopologySystems,zheng2024a}, sensory adaptation and kinetic proofreading \cite{Murugan2017TopologicallySystems}, population dynamics \cite{Knebel2020TopologicalCycles,Yoshida2021ChiralCycles}, and as building blocks for the design of synthetic biological or robotic systems \cite{Tang2021TopologySystems}. These theories could shed light on the fundamental question of why biological function is so robust, e.g. during development or when maintaining stable dynamics over long times, even in the presence of stochasticity or changing external conditions and stimuli.

}

In this review, we present different topological models that have been proposed in recent years for stochastic systems, along with a discussion of attractive properties of these models as well as the formalism used to describe them. This burgeoning field of study has also prompted the development of new theoretical and computational approaches, as well as the discovery of new physical phenomena. We survey these and discuss the relevant mathematical approaches. We close with a discussion of potential applications to biological and non-biological systems, which herald exciting future directions for the field. 

The review is organized as follows. In Section \ref{sec:background}, we give a brief overview of the development of topological phases initially for quantum matter, and later for classical systems in continuous, physical space. In Section \ref{sec:discreteclassical}, we describe recent discoveries of topological phases in discrete classical systems, where the dynamics take place in an abstract configuration space. These include stochastic dynamics in biochemical systems, and nonlinear deterministic dynamics in ecological systems. In the remainder of the review, we expand on the novel properties (Section \ref{sec:attractive}), applications \textcolor{black}{in molecular biology and synthetic systems with potential experimental realizations }(Section \ref{sec:applications}), and theoretical insights and tools (Section \ref{sec:tools}) that arise in stochastic systems in particular, although much of what we discuss also applies to population dynamics models.

\section{Background \label{sec:background}}

\subsection{Topological phases in quantum systems}
    
   Topology is concerned with the properties of objects that remain invariant under continuous deformations. While this can sound abstract, the study of topology has made powerful contributions to various physical problems, by distinguishing broad classes of materials and their resulting physical response through analysis of their fundamental properties. Its contribution to the field of quantum electronic systems has been striking, predicting which classes of materials exhibit currents or polarization localized to just the edges of the system \cite{Moore2010TheInsulators,Chiu2016ClassificationSymmetries,Stormer1999TheEffect,Cohen2019}. These edge currents are insensitive to local information such as impurities or defects. Due to this robust protection from disorder or noise, topological states have also been touted as candidates for fault-tolerant quantum computing \cite{Nayak2008Non-AbelianComputation}.
   
	{\color{black} Here, we specifically refer to the algebraic topological features of an object, which can be quantified using topological invariants that undergo discrete changes with the system parameters when one crosses from a topologically trivial phase to a topologically non-trivial phase. Indeed, several invariants have been identified for many different systems that satisfy the Schr{\"o}dinger equation $i\frac{d}{dt}\psi=\mc{H}\psi$. Fig. \ref{extfig:schematic} illustrates one example of a topological invariant, in this particular case a winding number. 
    
    Because topological models are typically composed of repeating motifs (unit cells) to form lattices, they can be Fourier transformed onto a periodic domain $k\in[0,2\pi)$. 
    In the Bloch sphere, the transformed Hamiltonian $\mc{H}(k)$ can wind around the origin as it traverses $k$ space, where the number of times it winds around the origin gives the winding number.  
This winding number is defined as
 \begin{equation}
     \gamma=\oint A_j(k)\cdot dk,\quad A_j(k)=-i\langle u_j(k)|\nabla_k|u_j(k)\rangle
 \end{equation}   
where $A_j(k)$ is known as the Berry connection and $u_j(k)$ is the eigenvector of $\mc{H}(k)$ for a given band $j$ \cite{Cohen2019}. This is also known as the geometric phase, or the Zak phase \cite{Obana2019TopologicalModel}. 

On the left of Fig.~\ref{extfig:schematic}, we depict systems with a nonzero winding number and hence in a topological phase. Due to the celebrated bulk-boundary correspondence \cite{Moore2010TheInsulators,Chiu2016ClassificationSymmetries,Thouless1982QuantizedPotential} between the lattice properties and its boundaries, this means that the original system (beige line) will host states on its boundary (red dots). Conversely on the right, systems which do not having a winding number are in the trivial phase, where there is no particular distinction between the system bulk  and its boundary (beige line).

Another example of a topological invariant is the Chern number, 
\begin{equation}
    c=\int\frac{d^2\vect{k}}{(2\pi)^2}\vect{\Omega}(\vect{k}), \quad \vect{\Omega}(\vect{k})=\nabla_{\vect{k}} \times\vect{A}(\vect{k}).\label{eq:Chernno}
\end{equation}
where bold script denotes a vector and the integral is taken over the two-dimensional Brillouin zone. The Chern number gives the transverse Hall conductance up to a factor of $e^2/\hbar$, which is such a precise and robust measurement that it has established a new standard for obtaining fundamental constants of nature \cite{Weis2011MetrologyEffect}. 

  More generally, the number and type of topological invariants that a material can have have been characterized just based on its dimensions and symmetry group \cite{Chiu2016ClassificationSymmetries}. Such systematic classifications were later extended to interacting systems \cite{Wang2014ClassificationDimensions,Tang2012InteractingPhases}. However, despite} these elegant and fundamental theoretical predictions, the accessibility and detection of topological states can be challenging in practice. Very clean, cold, and ordered samples are often required to observe these delicate quantum properties \cite{Moore2010TheInsulators,Stormer1999TheEffect}. 

\subsection{Topological phases in classical systems}
 
	 Topological approaches in biology have mostly been based on identifying \textcolor{black}{local} winding structures in biological data, including topological defects in tissues \cite{Shankar2022TopologicalMatter,DeCamp2015OrientationalNematics,Keber2014TopologyVesicles,Ardaseva2022TopologicalMatter,Saw2017TopologicalExtrusion} or knots in DNA strands \cite{Darcy2021DNALinks}, and more abstractly in any high-dimensional dataset through methods such as persistent homology \cite{Nicolau2011TopologySurvival,Reimann2017CliquesFunction,Taylor2015TopologicalNetworks,Giusti2015CliqueCorrelations,Giusti2016TwosData,Rocks2021HiddenFunctionality}. Other approaches have linked the geometric phase of a system with its dynamics or response, in the context of locomoting or rotating animals \cite{purcell1977life,Shapere1987Self-propulsionNumber,Shapere1989GaugeBodies} or stochastic chemical reactions \cite{Sinitsyn2007TheKinetics}.

  The last decade has seen diverse proposals for realizations of new phases of matter characterized by \textcolor{black}{global} topological invariants in classical platforms. These display edge localization and currents analogous to those in quantum systems, but can be probed in much less extreme conditions. Examples include mechanical lattices \cite{Mao2018MaxwellMechanics,Pedro2019TopologicalInteractions,Kane2014}, photonics \cite{Xiao2014SurfaceSystems,Benalcazar2017QuantizedInsulators,Yuce2019TopologicalModel,Pocock2018TopologicalEffects}, acoustics \cite{Ni_2017,Liu2017PseudospinsLattice}, and active matter \cite{Tang2021TopologySystems,DasbiswasE9031,Murugan2017TopologicallySystems}. 

  In all these examples of topological phases in classical systems, the dynamics of interest occur in actual physical space (e.g.~mechanical lattices, optical waveguides, microfluidic devices...) and are described by continuum mechanics, typically in a deterministic setup. In contrast, the dynamics of interest in biological systems often do not take place in physical space, but rather in an abstract and typically discrete configuration space representing concentrations of chemicals, conformational states of proteins, or population numbers of individuals. These dynamics are often stochastic and driven by chemical reactions or birth-death processes. Thus, a new approach is needed to identify and describe topological phases in these systems, which we describe in the following.


\begin{figure*}
      \includegraphics[width=\linewidth]{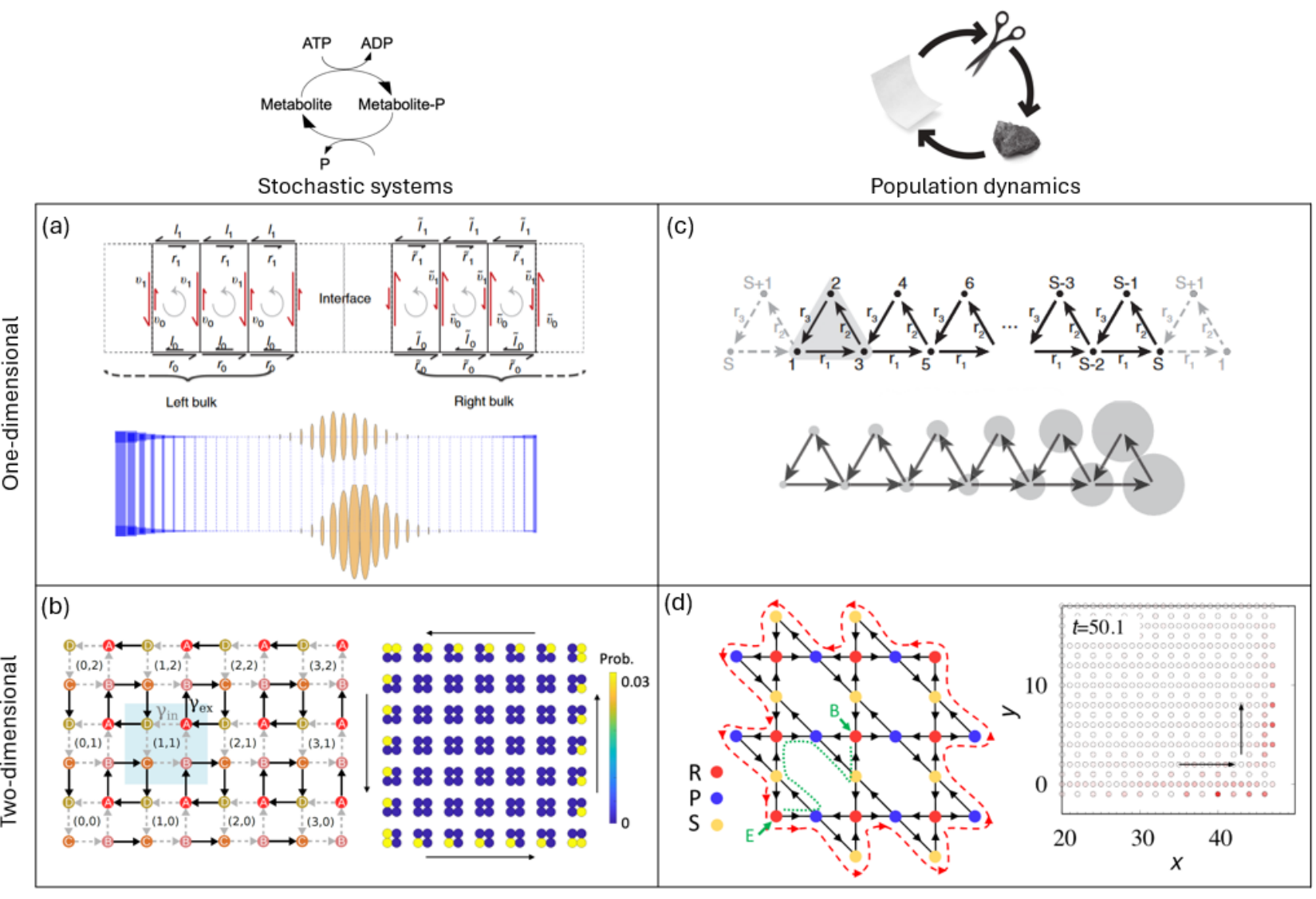}
      \caption{{Proposals for topological models in stochastic systems.} Non-equilibrium motifs (top) can be repeated to support a topological state.
      \emph{Left column}: Futile cycles, such as phosphorylation-dephosphorylation cycles, appear in various biological systems from protein synthesis to muscular contraction, metabolism and sensory systems \cite{hopfield1974kinetic,samoilov2005stochastic}. (a) A series of biochemical reactions form a 1D chain that displays edge states at the interface of two different topological invariants; adapted from Ref.~\citenum{Murugan2017TopologicallySystems}. (b) In a 2D lattice of interlinked futile cycles, a topological regime is found where probability accumulates at the edges of the system and a probability current with defined chirality spontaneously emerges, leading to system-spanning oscillations; adapted from Ref.~\citenum{Tang2021TopologySystems}.
      \emph{Right column}: The rock-paper-scissors interaction can model non-transitive evolutionary pressures in population dynamics.
      (c) In a 1D chain with such interactions, population can accumulate at one or the other end of the chain as a result of a topological phase transition; adapted from Ref.~\citenum{Knebel2020TopologicalCycles}. (d) In a 2D Kagome lattice, excesses in population density are transported along the edges of the lattice in a chiral manner, due to a topological effect; adapted with permission from Ref.~\citenum{Yoshida2021ChiralCycles}. \label{extfig:models}}
 \end{figure*}

\section{Topological phases in discrete classical systems \label{sec:discreteclassical}}

Repeated processes or motifs in configuration space, describing a basic reaction such as the conversion of biomolecules from one state to another, or the birth or death of individuals in a population, can be linked together to form an ordered network or lattice that allows for topological analysis. Most importantly, when many such motifs are linked together, it becomes possible to distinguish between the ``bulk'' of the lattice and its ``edges'', the latter arising e.g.~from a limiting supply of building blocks or other constraints in the reaction dynamics, that limit the extent of the configuration space. 

\textcolor{black}{For instance, this configuration space could be indexed by the numbers of repeated reactions, or the number of molecular subunits added or proteins transcribed. The repeated motifs would form the bulk, while edges are given by the lowest or highest numbers of the repeating index such as zero reactions or all possible reactions having taken place. Once such a configuration space has been mapped out, we can then apply onto the lattice a }
rigorous analysis of topological invariants, crucially extending models for equilibrium systems to a range of phenomena that is strongly non-equilibrium. They provide a description for the emergence of robust global responses from many basic building blocks (see Fig. \ref{extfig:models}).

A common example for a building block in the biochemical context is that of futile cycles \cite{hopfield1974kinetic,samoilov2005stochastic}; see Fig. \ref{extfig:models}, top left. Futile cycles consist of a series of chemical reactions that bring the molecule back to its starting point, a non-equilibrium process which dissipates energy and thus  requires the input of chemical fuel, typically provided by ATP or GTP in the biological context. Despite seeming wasteful (or futile, as their name suggests), such cycles can be found in protein synthesis, muscular contraction, metabolism, and sensory systems -- where their purpose across all these disparate systems remains unclear \cite{samoilov2005stochastic}. Intriguingly, the stochastic topological models that we describe below show that systems with futile cycles as building blocks tend to display topological states \cite{Murugan2017TopologicallySystems,DasbiswasE9031,Tang2021TopologySystems} . This suggests a fascinating route by which seemingly wasteful non-equilibrium motifs in biology can collectively foster robust global function.

A similar role to that of futile cycles has been found to be played by rock-paper-scissors motifs in ecological networks describing population dynamics \cite{Knebel2020TopologicalCycles,Yoshida2021ChiralCycles}; see Fig. \ref{extfig:models}, top right. These rock-paper-scissors motifs represent a non-transitive relation among three competing species: rock outcompetes scissors, which outcompetes paper, which outcompetes rock \cite{may1975nonlinear,szolnoki2014cyclic}. Non-transitive competition is widely prevalent in animals, plants, and microbes. As we describe below, when many of these rock-paper-scissors cycles are linked together, representing cases where a species plays a role in more than one such cycle within a larger ecosystem, topological states emerge at the scale of the whole ecosystem.

\subsection{Stochastic systems}

The stochastic dynamics of biochemical systems are typically described by continuous time Markov processes on a discrete state space, through a master equation of the form $\frac{d}{dt}\vect{p}=\mc{W}\vect{p}$, where $\vect{p}$ is a vector containing the probabilities $p_i$ of being in state $i$ (normalized as $\sum_i p_i=1$), and $\mc{W}$ is a transition matrix where $W_{ij}$ is the transition rate (probability per unit time) of the $j \to i$ transition \cite{schnakenberg1976network}. To enforce conservation of probability, the transition matrix must satisfy $W_{ij}>0$ for $i\neq j$ and $W_{jj}=-\sum_{i \neq j} W_{ij}$. These properties (together with the requirement that the directed graph corresponding to the network of transitions is strongly connected, so that there are no sinks or disconnected regions) guarantee that the transition matrix $\mc{W}$ has one zero eigenvalue, corresponding to the steady state $\vect{p}_\mathrm{ss}$ which satisfies $\frac{d}{dt}\vect{p}_\mathrm{ss}=0$, while all other eigenvalues have negative real part, representing exponentially decaying modes (which may be oscillatory or non-oscillatory depending on whether the imaginary part is nonzero or zero, respectively). This formalism is widely used to model conformational changes of proteins and protein complexes \cite{chodera2014markov}, the operation of molecular motors \cite{lipowsky2006molecular,kolomeisky2007molecular} and stochastic swimmers \cite{golestanian2008mechanical,chatzittofi2024entropy}, and the full stochastic dynamics of chemical reactions in finite systems  \cite{Qian2021StochasticBiology}.

\subsubsection{1D systems}

The first demonstration of topological states in stochastic systems was given for one-dimensional (1D) lattices in Ref.~\citenum{Murugan2017TopologicallySystems} and followed-up on shortly after in Ref.~\citenum{DasbiswasE9031}. A basic depiction of the model and results is shown in Fig.~\ref{extfig:models}(a). The basic lattice has a ``ladder'' shape, with unit cells containing two sites (top and bottom) that are replicated laterally along the horizontal direction to form a 1D bulk. Considering the transitions (both ways) among internal sites of the unit cell, and among top sites and bottom sites in neighboring cell, a given bulk is defined by six transition rates. The authors considered what occurs at the interface between two different lattices (defined by two different bulks, i.e.~two different sets of transition rates) when the two are linked together.

The authors showed that, for any given bulk, one can define a topological index, described \textcolor{black}{in Section~\ref{sec:formal}}. Crucially, when the topological indexes of the two bulks that are brought into contact do not coincide, localization (i.e.~accumulation of probability at steady state) is observed at the interface between the two. \textcolor{black}{Intuitively, when the topological invariants do not match, the left bulk and the right bulk have a tendency to create net currents pointing in opposite directions to each other, therefore causing accumulation either at the interface between the two bulks, or at opposite ends of the system.}

In summary, these seminal works demonstrated that topological considerations based purely on bulk properties can be used to predict global effects, occurring at the edges of a lattice or at the interfaces between two connected lattices. However, \textcolor{black}{because the net current between unit cells in a finite 1D system is necessarily zero at steady state}, the topological phenomena that could be observed were limited to static localization.

\subsubsection{2D systems}

These restrictions were lifted by us in Ref.~\citenum{Tang2021TopologySystems}, which provided the first realization of topological states in two-dimensional (2D) stochastic systems. In 2D systems, besides static polarization, topologically-protected chiral currents along the edges of the system, analogous to those seen in the quantum Hall effect \cite{Cohen2019}, become possible.

In the minimal model, see Fig.~\ref{extfig:models}(b), the unit cell has four sites, labelled A--D, and is replicated along two directions forming a 2D lattice (note that a three-site model can also be defined \cite{Tang2021TopologySystems}, but is less amenable to analysis). The four sites of a unit cell are connected by four transitions forming an ``internal'' futile cycle. In the bulk of the system, four neighboring unit cells are also connected by four transitions forming an ``external'' futile cycle. The minimal model, with unidirectional transitions, thus has eight transition rates in total (which become sixteen transition rates if bidirectional transitions are allowed). However, one can further impose rotational symmetry and make all internal transitions equal to each other, and similarly for external transitions, so that only two transition rates (internal and external) remain and define the bulk behavior.

Importantly, when (i) internal and external cycles have opposite chirality and (ii) internal transitions are much slower than external ones, probability accumulates at the edges of the system, and a probability current is observed along these edges, with the same chirality as that of the internal cycles. Thus, global oscillations are observed at a system-spanning scale. Further analysis, described \textcolor{black}{in Section~\ref{sec:formal}}, shows that this effect is topological and thus the edge currents are protected against disorder and deformations of the system boundary.



\subsection{Population dynamics \label{sec:popdyn}}

Population dynamics models are typically described at the mean field level, with non-linear equations that describe the number, concentration, or fraction of individuals of different species in a well-mixed (spatially homogeneous) ecosystem. The simplest and most commonly studied description for the competition between different species is the Lotka-Volterra Equation \cite{goel1971volterra}.  For the particular case of a zero-sum game in which the total population is conserved, and a decrease in the population of one species is balanced by an increase in the population of another, the deterministic dynamics are described by the Antsymmetric Lotka-Volterra Equation (ALVE), which can be written as $\frac{d}{dt} x_i = x_i \sum_j A_{ij} x_j$, where $x_i$ is the population of species $i$, and $\mc{A}$ is an antisymmetric interaction matrix with components $A_{ij} = -A_{ji}$ that define the transfer of population between species $i$ and $j$ \cite{goel1971volterra,knebel2015evolutionary}. The total population is conserved, and without loss of generality can be normalized to $\sum_i x_i = 1$. The ALVE also appears in various contexts other than population dynamics, such as game theory, Bose-Einstein condensation, plasma physics, and chemical kinetics \cite{knebel2015evolutionary}. Note that, while bearing some similarities to the master equation described above, as both describe the dynamics of a conserved quantity that is transferred across discrete sites, the ALVE  describes very different dynamics: first, it is nonlinear (as it describes a many-body interacting system at the mean field level) while the master equation is linear; second, the matrix $\mc{A}$ is antisymmetric while $\mc{W}$ has no such constraint, although it does have other constraints. 

\subsubsection{1D systems}

As in the case of biochemical systems, the first demonstration of topological states in population dynamics was given for a 1D lattice in Ref.~\citenum{Knebel2020TopologicalCycles}. The authors linked together many rock-paper-scissors cycles into a long chain, see Fig.~\ref{extfig:models}(c). The bulk of the system is thus defined by the three interaction strengths that define the cycle. By construction, in each cycle there is always one `isolated' species that only interacts internally with the other two species in the cycle, and two `connected' species that also take part in the neighboring cycle, and thus interact with four species in total.  The authors found that, depending on the skewness of the cycle, determined by the ratio of interaction strengths between the isolated and the two connected species, population would accumulate on either the left or the right end of the chain. Thus, similarly to the 1D biochemical systems, static polarization is observed at steady state. Importantly, polarization was robust to initial conditions and to small disorder in the system parameters, suggesting a topological origin of the transition. This was confirmed by a mapping of the interaction matrix to a Hermitian Hamiltonian that could be analyzed using standard tools of quantum topology, described further below. We note as well that the effect of deviations from ideal rock-paper-scissors interactions in the 1D chain was recently studied in Ref.~\citenum{yoshida2022non}.

\subsubsection{2D and 3D systems}

In Ref.~\citenum{Yoshida2021ChiralCycles}, a 2D Kagome lattice of rock-paper-scissors cycles was studied, see Fig.~\ref{extfig:models}(d). All interactions were chosen to be identical, so that the system is essentially parameter-free. The authors found that, when the system is initialized with additional population in one of the edge sites of the lattice, this additional population propagates along the edges of the lattice, with the same chirality as that of the rock-paper-scissors cycles. As in 2D biochemical systems, this edge current \textcolor{black}{seemed} to be topologically protected, \textcolor{black}{as it was found to be robust to deformations or defects in}  the shape of the boundaries. By mapping the interaction matrix to a Hermitian Hamilonian in a similar way as for the 1D chain, they could show that the edge current is induced by a nontrivial topological index (\textcolor{black}{the Chern number, see Eq. \ref{eq:Chernno}}).

By stacking several of these 2D Kagome lattices on top of each other along a third ($z$) axis, a 3D lattice of rock-paper-scissors cycles was constructed in Ref.~\citenum{umer2022topologically}. Similarly to the 2D case, chiral propagation of population density was observed in the dynamics projected onto the $xy$ plane, implying density propagation along the faces of the system. No significant dynamics were reported along the $z$ dimension. The chiral propagation was shown to be robust and to have a topological origin, through a similar mapping as for the 1D and 2D cases.

\subsection{Formalism and mapping to quantum systems \label{sec:formal}}

Topological invariants in stochastic systems can be identified using results developed in quantum systems. 
In Ref.~\citenum{Murugan2017TopologicallySystems}, a topological index was identified from the winding number in Fourier space of the phase associated to the determinant of the tilted current matrix associated to the bulk. In follow-up work, this index was further related to the topological index of the Hermitian Hamiltonian of an associated supersymmetric quantum system \cite{DasbiswasE9031}.

A more direct method that flexibly maps many systems described by the master equation $\frac{d}{dt}\vect{p}=\mc{W}\vect{p}$ to a quantum Hamiltonian was used in Ref.~\citenum{Tang2021TopologySystems}, although the resulting Hamiltonian in this case is typically non-Hermitian \cite{Okuma2023Non-HermitianReview,Kawabata2019SymmetryPhysics}. Namely, the similarity of the master equation to the Schrödinger equation $i\frac{d}{dt}\psi=\mc{H}\psi$ is exploited, as they are formally identical up to the constant $i$. In quantum systems, topological invariants such as the Chern number are typically calculated from the eigenvectors of the Hamiltonian $\mc{H}$\textcolor{black}{, see Eq. \ref{eq:Chernno}
}. When mapping a master equation to a corresponding Schrödinger equation, the eigenvectors of the transition matrix $\mc{W}$ will be identical to those of $\mc{H}$, in the case of periodic boundary conditions. Hence, the same calculation that verifies if the quantum system is topological using its eigenvectors, can be performed for $\mc{W}$ from a stochastic system. Using the bulk-boundary correspondence, a topological invariant that characterizes the bulk will predict a boundary response for the system. Further works have shown that this bulk-boundary correspondence has additional constraints in stochastic systems as compared to quantum ones \cite{Nelson2024NonreciprocitySystems,Sawada2024RoleProcesses,Sawada2024Bulk-BoundaryProcesses}, which is discussed below in Section \ref{sec:tools}.

A similar approach can be used in nonlinear models of population dynamics described by the ALVE. In particular, when the ALVE is linearized around a uniform steady state solution, with deviations $\delta x_i = x_i - 1/N$, where $N$ is the number of species in the lattice, the equation can be written as  $i\frac{d}{dt} \delta \vect{x} =\mc{H} \delta \vect{x}$, which has the same form as the Schrödinger equation with a Hamiltonian $\mc{H}=i \mc{A} / N$ \cite{Yoshida2021ChiralCycles}. In this case, because the interaction matrix $\mc{A}$ is antisymmetric, the Hamiltonian $\mc{H}$ is Hermitian, and the usual tools of Hermitian quantum topology can be used \cite{Knebel2020TopologicalCycles,Yoshida2021ChiralCycles,umer2022topologically}.

\begin{figure}
      \includegraphics[width=\linewidth]{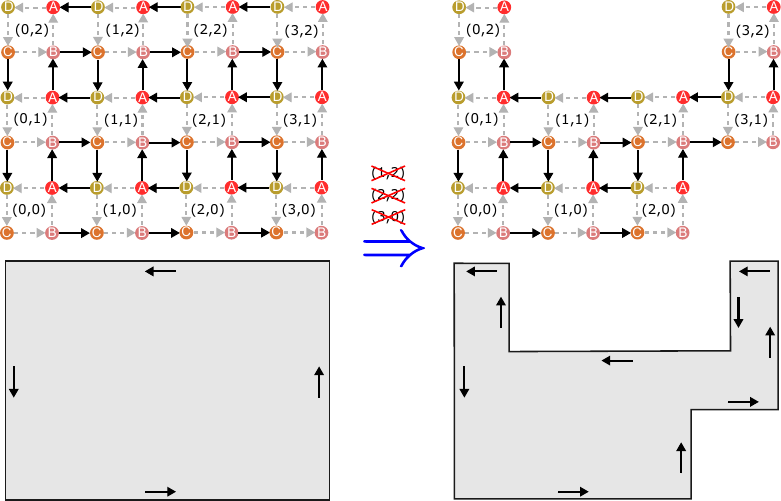}
      \caption{{Topological protection ensures robustness of the edge current to obstacles or missing components.} When certain components of the system are missing or inaccessible (states with red crosses), the edge current will simply go around them to maintain the largest available phase space, when in the topological phase [$\gex\gg\gin$, see Fig.~\ref{extfig:models}(b)]. This robustness of the edge state can shed light on how biological systems flexibly pivot in the presence of changing conditions or external stimuli. \label{extfig:robust}}
 \end{figure}

\section{Attractive features and useful properties \label{sec:attractive}}

{\color{black}In the remaining sections of the review, we restrict our focus to stochastic systems, for which there have been a greater number of follow-up works exploring further methods of analysis and potential applications. In particular, while periodic lattices that allow the definition of a bulk and a boundary arise naturally in stochastic systems, the appearance of such lattice structures in ecological networks is less obvious (additionally, realistic population dynamic models do not generally have the conservation laws that allow their reformulation as an ALVE). Nevertheless, many of the conclusions drawn in the following may also be applied to such population dynamics models. }

\subsection{Robustness}
      
Topological states are known to be robust to deformations of the edge, and hence to impurities or disorder. The global response in such stochastic systems therefore inherits the useful property of topological protection that provides robustness to missing components or obstacles, see Fig. \ref{extfig:robust} \cite{Tang2021TopologySystems,zheng2024a}. This behavior can shed light on how biological function remains so robust, e.g. during development or when maintaining stable dynamics over long times, even in the presence of stochasticity or changing external conditions and stimuli. 
 
Numerical simulations have been conducted to test the robustness of these topological models to random noise or perturbations, which are common in biological systems. By systematically adding random fluctuations to transition rates, these simulations have shown that weak disorder does not destroy the topological phase \cite{Murugan2017TopologicallySystems,zheng2024a}.
    
\subsection{Dimensional reduction of responses and dynamics} 

The reduction of the full system response to a much lower dimensional description has been observed in many complex biological systems, where the system dynamics or behavior is confined to a much smaller subset of the phase space \cite{rigotti2013importance, tang2019effective, stephens2008dimensionality}. This is exemplified in computational models of memory, that describe specific attractor states which represent persistent memories \cite{chaudhuri2016computational, hopfield1982neural}.  Another example is that of long oscillations, such as the circadian rhythm, which are crucial for the regulation of many processes such as metabolism and replication \cite{liao2021circadian, puszynska2017switching}.  Confining the dynamics onto the lower dimensional boundary of a much larger network, topological models in stochastic systems offer a route to achieve this ubiquitous phenomenon. 

As topological models demonstrate how to achieve dimensional reduction in a precise manner, stochastic versions provide the necessary and sufficient conditions under which biological networks, notoriously large and messy, can reduce their activity to just a submanifold of their full phase space. 
Moreover, because edge states decay exponentially away from the system edge with typically short decay lengths \cite{Tang2021TopologySystems}, they can persist even in small systems. Short decay lengths also make it possible for domains with distinct topological invariants to co-exist on the same network, allowing for localized states or currents at the boundary between subnetworks or subregions. 

\subsection{Small motifs that can be scalable}
  
Topological models typically consist of lattices, which provide a versatile description as they consist of minimal biochemical motifs describing just a few interlinked reactions (see Fig. \ref{extfig:models}). These underlying building blocks are repeated across the network and support the emergence of robust global responses. As the same motifs are repeated, the resulting responses are governed by just a few parameters. Meanwhile, the repeated motifs can scale up towards large networks and systems, providing a flexible design architecture.

\section{Applications to biology and synthetic systems \label{sec:applications}}

The discrete nature of molecules and proteins, which moreover often exist only in small copy number, naturally lends itself to discrete stochastic descriptions. In these systems, the stochastic nature of discrete events can yield new physics, such as extinction events in gene regulation \cite{Wolynes2010GeneContext}. Topological modeling of this new physics can provide insights about robustness in chemical and biological systems, that can be further extended to synthetic systems. 

\begin{figure*}
      \includegraphics[width=0.9\linewidth]{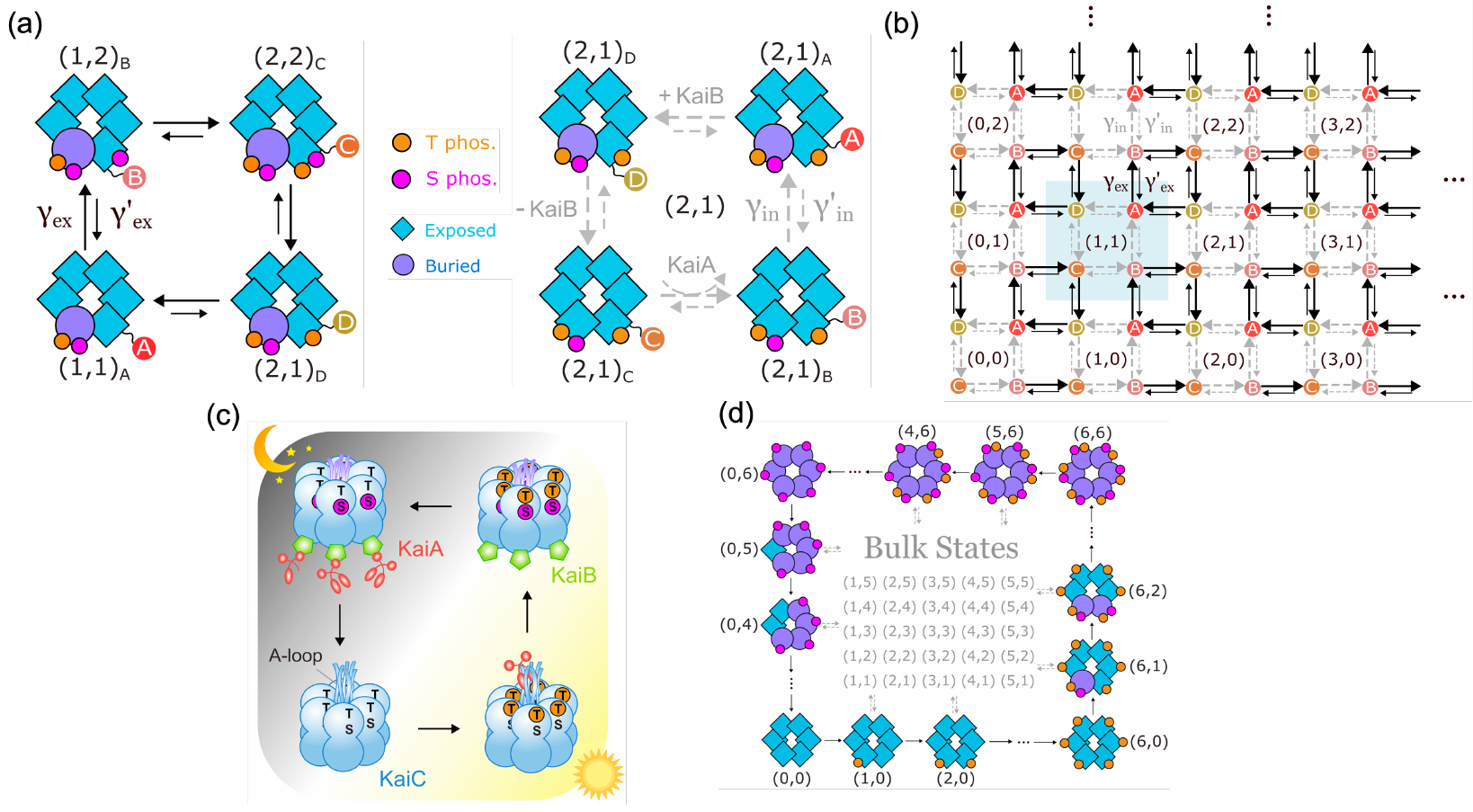}
      \caption{{Topological model for emergent oscillations.} (a) Macromolecules exhibit a large space of possible conformations, here illustrated with the KaiC hexamer that governs the circadian rhythm of cyanobacteria. Based on observations of autophosphorylation in the literature \cite{Brettschneider2010AClock.,Kageyama2006CyanobacterialVitro,Lin2014MixturesClock}, it is thought that monomers undergo phosphorylation and dephosphorylation cycles (black arrows, $\gex$), as well as conformational changes between exposed and buried configurations, or binding with other proteins such as KaiA and KaiB (grey arrows, $\gin$). (b) These cycles can be laid out following the model of Fig.~\ref{extfig:models}(b), with reversed rates included in this case. (c) KaiABC exhibits oscillations via a concerted global cycle of phosphorylation and dephosphorylation. During the day, all six KaiC monomers get phosphorylated at the T-sites, and then at the S-sites. This phosphorylation phase is promoted by interaction with KaiA molecules \cite{xu2003cyanobacterial}. By night, fully phosphorylated KaiC binds to KaiB, which sequesters KaiA from the solution. In the absence of KaiA, all the T-sites get dephosphorylated, followed by the S-sites \cite{rust2007ordered}. 
      Since individual monomers can independently phosphorylate  \cite{Brettschneider2010AClock.}, it is unclear why they would perform a concerted phosphorylation cycle that is robust.  (d) A possible solution lies in the topological phase of the model, in which a global cycle emerges that recapitulates the experimentally observed phosphorylation sequence. Adapted from Ref.~\citenum{zheng2024a}.  \label{extfig:kai}}
 \end{figure*}
\subsection{Molecular biology}

Non-equilibrium motifs abound in many different biochemical systems, making the formalism described here applicable to macromolecular or cellular dynamics.  It is thus desirable to investigate how topology can be realized in a biological system, given the many attractive properties of topology such as its robust response. 

\subsubsection{Long timescales: the circadian rhythm}

 A detailed case study of a topological mechanism in a concrete biological system is that of the KaiABC system for the circadian clock of cyanobacteria \cite{ishiura1998expression,nakajima2005reconstitution}. As highlighted in  Ref.~\citenum{Tang2021TopologySystems}, topological edge currents  in Fig. \ref{extfig:models}(b) naturally reproduce the kinetic ordering of KaiC phosphorylation cycles, although the detailed biophysical mechanisms remained abstract in this work. These missing mechanisms were made concrete \textcolor{black}{(see Fig.~\ref{extfig:kai})} in Ref.~\citenum{zheng2024a}, which also characterized the performance and efficiency of the resulting clock. In particular, it was shown that the key biochemical mechanisms \cite{kim2008day,chang2011flexibility,tseng2014cooperative,hong2018molecular} interact via a separation of time-scales to produce the resulting edge currents, and provide the necessary and sufficient conditions for the observed oscillatory cycle \cite{zheng2024a}. An analysis of the parameter space showed how a variation of the transition rates (e.g. by changing ATP concentration \cite{phong2013robust} or using a mutant \cite{qin2010intermolecular}) affects the coherence and dissipation of the oscillation, by tuning the system into and out of the topological transition.  This yields key insights into an important regulatory system that had until now required rather complicated models, especially to reproduce the observed kinetic ordering of T and S phosphorylation \cite{van2007allosteric,paijmans2017thermodynamically,smolen2001modeling}.

On the theoretical level, Refs.~\citenum{Tang2021TopologySystems} and \citenum{zheng2024a} characterized the coherence of the resulting cycle, showing that it satisfies theoretical bounds \cite{barato2017coherence} for the most coherent oscillator, equivalent to that of a unicyclic network, without the fine-tuning needed for a unicycle. The topological model shows high coherence compared to other available models, while producing the global day-night cycle with unusually few free parameters. In addition, the coherence and energetic cost of the oscillation were explored using tools from non-equilibrium stochastic thermodynamics, to reveal an efficient regime where coherence increases while cost simultaneously decreases. {\color{black}This finding provides an experimental prediction where one could test if decreasing ATP concentration increases ADP consumption, which is the opposite of what is usually expected. 

Further, cryoEM could be used to distinguish between this topological model and the typical MWC paradigm, since the former predicts new conformational distributions. Specifically, the topological model should show a wide distribution of different A-loop conformational patterns, as opposed to a narrow distribution that is expected from MWC models. One would also expect that mutants mimicking phosphorylated S-sites (T-sites) would correlate with hexamers having mostly buried (exposed) A-loops in the topological model, whereas such correlations would not be expected otherwise. More broadly, a topological model is able to explain long-standing puzzles in biology, such as how dimensional reduction is achieved in a robust and flexible manner to produce emergent robust oscillations. }

\subsubsection{Sensory adaptation}

In Ref.~\citenum{Murugan2017TopologicallySystems}, the authors used their 1D topological model [Fig. \ref{extfig:models}(a)] to describe sensory adaption, in particular how \textit{E. coli} senses chemical concentrations to navigate in response to chemical gradients \cite{barkai1997, tu2013}. Here, the activity of the bacterial sensing complex is modified by its methylation level according to a sigmoid function. More specifically, there is low activity at low methylation, high activity at high methylation, and a narrow region in between where activity changes fast. The sensing complex tends to have a methylation level where activity changes the fastest, which ensures maximal sensitivity to chemoattractant.

The authors propose that their network, with a region of low methylation and activity and a region of high methylation and activity, can be described by a topological model with two different topological phases. The steady state is localized at the domain wall between the two phases. This ensures that the system is localized in the region where activity changes the fastest, allowing for rapid adaptation to environmental changes, while the topological origin for this mechanism makes the system robust against random perturbations.

{\color{black}
\subsubsection{Gene transcription}

Topological invariants have also been used to predict different dynamical regimes in a simple gene switch \cite{schultz2007} and demonstrate the non-equilibrium phase transitions between these regimes. Here, topology provides new insights where a simple energy landscape paradigm fails: that multistable steady states can lie on the edges of a topological model. In particular, topological analysis reveals a new oscillatory regime in addition to previously recognized bistable and monostable phases \cite{nelson2025topology}. The authors show how local winding numbers predict the steady-state locations in the bistable and monostable phases, while a flux analysis quantifies the respective strengths of steady-state peaks. 

This work predicts different transcription regimes (e.g., oscillations vs random switches) that can be distinguished experimentally. Further, it uncovers different mechanisms that lead to multistability or oscillations, that can be useful to quantify existing observations in synthetic gene circuits \cite{zhu2022}. Notably, this work develops new methods for systems without a repeated lattice structure or that have disorder, extending the suite of biological networks that can be analyzed using topology \cite{nelson2025topology}.}

\subsubsection{Dynamical instability of microtubules}

Using a 2D Kagome model with edge-localized currents \cite{Tang2021TopologySystems}, it was shown that similar dynamics to the growth and shrinkage of microtubules can arise from topologically-protected chiral edge currents. Specifically, the topological model predicts stable phases of growth followed by sporadic phases of shrinkage using only three parameters. While further work is needed to probe the specific biophysical features of the system, the model is able to successfully capture the wide range of catastrophe lengths and merits further investigation.

\begin{figure}
      \includegraphics[width=\linewidth]{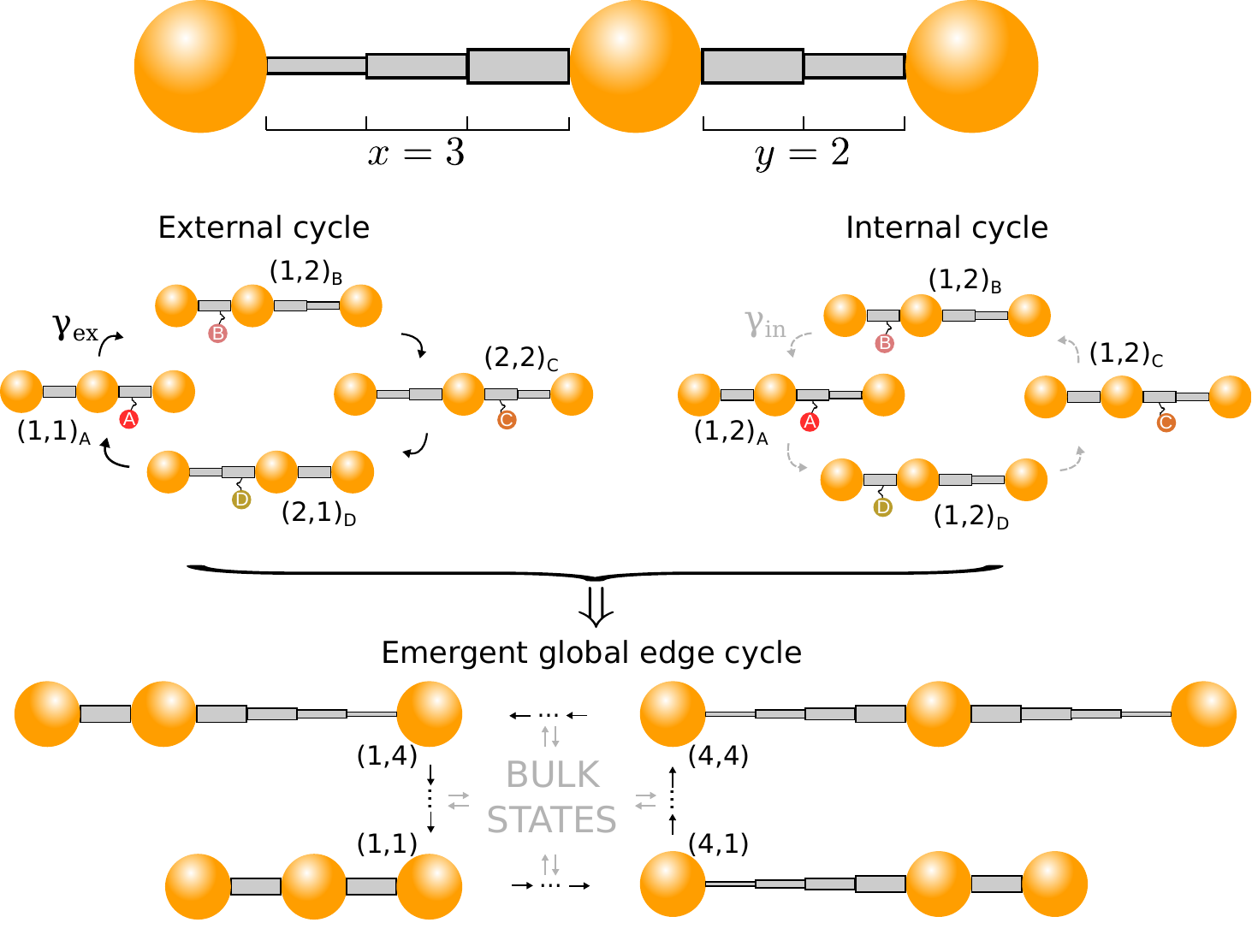}
      \caption{{A topological design for a microswimmer can flexibly navigate around malfunctions or obstacles.} The velocity of a 3-sphere microswimmer is proportional to the area it encloses in shape space \cite{Shapere1987Self-propulsionNumber,golestanian2008mechanical}. Topologically-protected edge cycles enclose all the available shape space and thus maximize the velocity of the microswimmer, even in the presence of component malfunctions or obstacles. Adapted from Ref.~\citenum{Tang2021TopologySystems}. \label{extfig:swimmer}}
 \end{figure}
{\color{black}
\subsection{Synthetic systems}
\subsubsection{Microfluidic networks}

Due to their tunability and ease of control, microfluidic devices have become natural platforms for the design and realization of different topological states \cite{shankar2020topological}. This includes the patterning of a Lieb lattice in which an active liquid can form a topologically protected sound mode \cite{Souslov2017TopologicalMetamaterials}, or the construction of chiral topological flows \cite{Copar2020MicrofluidicFlows}.

Active particles can also be guided to accumulate at corners through a second-order skin effect \cite{Palacios2021GuidedEffect}, an example of higher-order topological phases, which cause the accumulation of particles at the zero-dimensional corners of a two-dimensional system. 
More generally, in Ref.~\citenum{Palacios2021GuidedEffect}, a microfluidic device with open chambers connected to each other by narrow connections of various shapes and widths resulted in the active particles jumping stochastically between chambers, thus providing a direct mapping to the systems described in the present review. Microfluidic systems containing active particles therefore offer a versatile testbed for lattices or geometries that support new topological phases and edge states.

\subsubsection{Robust microswimmer design}}

The 2D topological model such as the one in Fig. \ref{extfig:models}(b) provides a novel design for low-Reynolds number microswimmers \cite{purcell1977life,Shapere1987Self-propulsionNumber}, as depicted in Fig. \ref{extfig:swimmer} for the example of a stochastic three-sphere swimmer \cite{golestanian2008mechanical,chatzittofi2024entropy}. By modeling the extension of the left and right arms of the swimmer as the two dimensions of our model, the topological edge state will naturally enclose the largest possible area in shape space, which is directly proportional to the self-propulsion velocity of the microswimmer \cite{Shapere1987Self-propulsionNumber,golestanian2008mechanical}. Notably, if either of the swimmer arms encounters a malfunction that e.g.~limits its extension, the edge current would simply move around this obstacle to maintain the largest available phase space. Such models can be useful for the design of synthetic microswimmers or macroscopic robots in complex environments \cite{Hatton2013GeometricMedium} that exhibit flexible behavior. 

\section{New theoretical tools and insights \label{sec:tools}}

The discovery of topological phases in stochastic systems has prompted the development of new theoretical tools to analyze them. This has revealed interesting new properties that are distinct from quantum and other topological systems, opening up new fields for investigation and study.

\subsection{Necessity of non-Hermiticity and out-of-equilibriumness}
  
\begin{figure}
    \centering
    \includegraphics{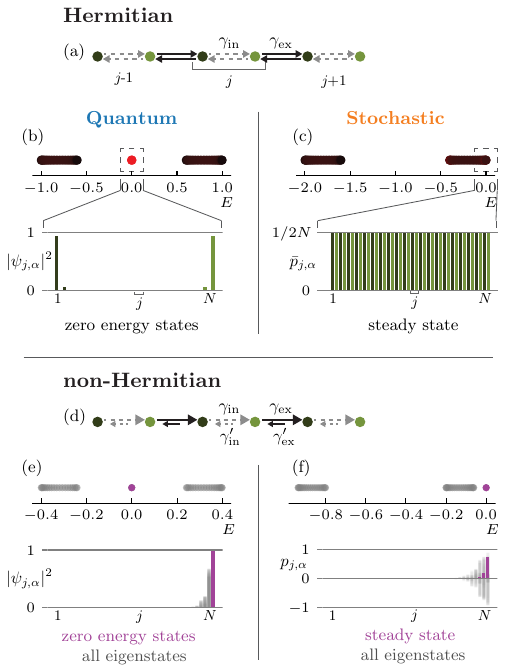}
  \caption{ {A non-Hermitian transition matrix is needed for topological edge effects in stochastic systems.} Models that are fully Hermitian (a) show non-trivial steady states in quantum systems (b), but only show homogeneous ones in the stochastic counterpart (c). This is due to the probability-conserving diagonal term that stochastic systems have in addition to the off-diagonal term shared with quantum systems \cite{schnakenberg1976network}. When non-Hermiticity is present (d), edge localization is possible in both quantum (e) and stochastic (f) systems. This behavior is shown here in 1D chains but persists in higher dimensions \cite{Nelson2024NonreciprocitySystems}. Adapted from Ref.~\citenum{Nelson2024NonreciprocitySystems}. 
    }
    \label{fig:nonreciprocal}
\end{figure}

While there is strong motivation to realize topological states in stochastic systems as described above, general principles that govern \textcolor{black}{topological features in stochastic systems, such as the edge responses necessary for robust function,} remain lacking. Recent theoretical work has addressed this question and demonstrated that the only way for stochastic systems to exhibit a localized response is to have a non-Hermitian transition matrix $\mc{W}$, see Fig. \ref{fig:nonreciprocal} \cite{Nelson2024NonreciprocitySystems}. Note that Hermitian transition matrices always describe equilibrium systems with uniform steady states, whereas non-Hermitian ones can (but not necessarily do) describe out-of-equilibrium ones. This is in line with other phenomena that have no counterparts in equilibrium systems \cite{hopfield1974kinetic,qian2000pumped,Shankar2022TopologicalMatter}.

Interestingly, this non-Hermiticity condition is in sharp contrast with quantum systems, where Hermitian topological invariants are known to cause edge responses. This work establishes a strict condition for stochastic topological localization. 
Further, Ref.~\citenum{Nelson2024NonreciprocitySystems} showed that the edge response is markedly different in stochastic systems compared to quantum systems. They also demonstrate a novel mechanism by which non-Hermiticity engenders the steady-state current on the edge \cite{Nelson2024NonreciprocitySystems}. These findings reveal surprising differences between quantum and stochastic edge responses despite both systems sharing the same topological invariant.

\subsection{Topological invariants and relaxation time}

The results above suggest the need for novel topological invariants designed for stochastic systems and their steady states, that signal more precisely the emergence of edge states. Such new tools have been developed for 1D stochastic systems \cite{Sawada2024RoleProcesses,Sawada2024Bulk-BoundaryProcesses}. Specifically, the authors introduced a winding number using a scale-transformed transition-rate matrix in periodic boundary conditions, which predicts the existence of a spectral gap between the steady state and other states \cite{Sawada2024RoleProcesses}. Further, this gap characterizes the relaxation time of the system towards the steady state: when the system is in the topological phase, the relaxation time scales with system size $O(N)$. In contrast, the trivial phase is expected to be gapless and have a relaxation time that scales as $O(N^2)$. Subsequently, the authors extended these results from ergodic systems also to non-ergodic ones, and developed a bulk-boundary correspondence between the winding number under periodic boundary conditions and the number of localized steady states in semi-infinite boundary conditions \cite{Sawada2024Bulk-BoundaryProcesses}.  

\subsection{Transfer matrix approach}

The key quantities that define a topological state with chiral edge currents in a stochastic systems are the steady state accumulation of probability at the edges of the system, and the steady state edge current, i.e.~how fast and in which direction this probability current flows along the edges of the system at steady state, see Fig.~\ref{extfig:models}(b). In Ref.~\citenum{Tang2021TopologySystems}, a novel method to obtain these quantities was proposed, based on a transfer matrix approach. The key to the approach is to relate the steady state probabilities of neighboring cells to each other. In particular, for the model in Fig.~\ref{extfig:models}(b), one can obtain a relation $\vect{P}_n = \mc{M} \vect{P}_{n-1}$, where $\vect{P}_n$ is a four-component vector containing the probabilities at the four sites of the unit cell that is a distance $n$ away from the edge of the system, and $\mc{M}$ is a $4 \times 4$ transfer matrix, independent of $n$. The eigenvalues and eigenvectors of $\mc{M}$ then contain all the necessary information about the edge states. In particular, $\mc{M}$ was found to have two eigenvalues equal to one, corresponding to a uniform steady state of the bulk of the system, and two eigenvalues $\alpha \in [0,1)$ and $1/\alpha$, corresponding to edge modes located at opposite edges of the system. From the eigenvector associated to $\alpha$, the probability accumulation and edge current could then be explicitly calculated \cite{Tang2021TopologySystems}. Additionally, $\alpha$ corresponds to the decay length of the edge mode, which is damped by a factor $\alpha^n$ in a unit cell a distance $n$ away from the edge.

\subsection{Topological classification using symmetries}

The classification of topological phases has its origins in results from random matrix theory \cite{Schnyder2008ClassificationDimensions,Ryu2010TopologicalHierarchy,Mehta1967RandomLevels}. This is based on the study of modes that remain gapless under arbitrary perturbations of the Hamiltonian that preserve the characteristic symmetries, including disorder. It was found that the results depend only on the system's symmetry, and Altland and Zirnbauer identified ten groups in total, an exhaustive list \cite{Wigner1955CharacteristicDimensions,Dyson1962TheMechanics,Altland1997NonstandardStructures}. Specifically, they determined which surface modes can completely evade Anderson localization from random impurities, depending on the system dimension and symmetry class \cite{Schnyder2008ClassificationDimensions,Ryu2010TopologicalHierarchy}. 

The ten Altland-Zirnbauer groups form a classification of Hermitian matrices and their possible topological phases. Many of the early proposals for topological phases in discrete classical systems indeed map their systems into Hermitian Hamiltonians and use the topological properties associated to the relevant symmetry class and dimension \cite{DasbiswasE9031,Knebel2020TopologicalCycles,Yoshida2021ChiralCycles,umer2022topologically}, employing the bulk-boundary correspondence to predict the appearance of edge states. However, stochastic systems governed by a master equation require an explicitly non-Hermitian description to exhibit surface states, when directly mapped to the Schr{\"o}dinger equation \cite{Nelson2024NonreciprocitySystems}; see Fig. \ref{fig:nonreciprocal}. There exist non-Hermitian features with no Hermitian counterpart, including the skin effect and an enhanced boundary sensitivity to random perturbations \cite{Okuma2023Non-HermitianReview}. While some topological classifications of non-Hermitian Hamiltonians have been recently proposed \cite{Kawabata2019SymmetryPhysics}, the relation of non-Hermitian topology to stochastic systems and the relevant symmetries for biological systems are questions that remain open.

{\color{black}

\subsection{Stochastic thermodynamics}

An interesting alternative approach using thermodynamic properties and the stoichiometric matrix associated to the transition matrix has been suggested, that can also predict the topological transition \cite{Mehta2022ThermodynamicSystems}. Separately, analysis of the entropy production in a topological model has revealed a regime of high efficiency, in which there is less dissipation despite greater coherence \cite{zheng2024a}. Stochastic thermodynamics and the associated fluctuation theorems \cite{seifert2012stochastic} can also be used to infer the non-equilibrium driving forces (entropy production) from observable trajectories, sometimes with close connection to the underlying topology \cite{mahault2022topological}.

\subsection{Dynamical phases in interacting systems}

Besides the population dynamics models described in Section \ref{sec:popdyn}, which corresponded to the mean field level of an interacting system, most of the works discussed so far concerned noninteracting systems, i.e.~effectively single-particle models in one- or two-dimensional lattices. One particularly interesting direction for future work is the application of topological tools to interacting, many-particle stochastic systems. Inspired by biological systems under resource constraints, Ref.~\citenum{Tang2021TopologySystems} considered two coupled systems with shared boundaries, resulting in (anti)synchronized edge states that were studied numerically.

Another recent work \cite{PhysRevE.109.L032105} explored a paradigmatic example of interacting dynamics, namely the symmetric simple exclusion process, and showed how symmetry-protected topological phases can be used to identify distinct dynamical phases with regards to the types of rare trajectories that emerge in the system. Future work in this direction could explore genuinely non-equilibrium interacting processes, such as asymmetric exclusion processes, and apply the same tools to interacting generalizations of systems in which even the single particle dynamics is topological, such as those described above.
}

\section{Outlook}

\textcolor{black}{While we have highlighted possible experimental realizations in synthetic gene circuits or single-molecule measurements, future work remains to be done in the verification of existing topological models for settings of biological relevance. On the theoretical side, we reviewed several developments and advances in the previous section. Still, we look forward to the development of new models that could exhibit as yet unknown features that are exclusive to stochastic systems and have no quantum counterpart. Lastly, while there has been initial progress in developing tools amenable to systems with disorder or heterogeneity \cite{nelson2025topology}, more work needs to be done to analyze the robustness of existing models to other forms of disorder or constraints imposed by real-world scenarios. }

{\color{black}
It should be noted that most of the works described here dealt with one- or two-dimensional systems. Most biomolecular and ecological phenomena, on the other hand, take place in much higher-dimensional spaces. Future work should therefore explore topological phenomena in high-dimensional systems and, relatedly, the higher-order topological phenomena that will be needed to localize the system's dynamics to much lower dimensional manifolds. While significant progress has been made in understanding higher-order topological phases in quantum systems \cite{xie2021higher,Yang_2024}, very little is known about their emergence in classical stochastic systems.
}

More broadly, this emerging field renders the powerful tools of topology and its robust dimensional reduction despite random perturbations, more accessible. Going forward, this research can lead to design prescriptions for targeted responses in synthetic biology, active matter, ecology, and robotics. This provides tools for building systems with stable function in the heterogeneous, non-equilibrium regimes which are the norm in everyday life.



\begin{acknowledgments}
We thank Aleksandra Nelson for helpful comments. E.T. acknowledges support from the NSF Center for Theoretical Biological Physics (PHY-2019745) and the NSF CAREER Award (DMR-2238667).
\end{acknowledgments}

\end{document}